\documentstyle[12pt,fleqn]{article}
\def\nel{N_{\rm el}}

\def\beeq{\begin{equation}}
\def\eneq{\end{equation}}
\def\beeqa{\begin{eqnarray}}
\def\eneqa{\end{eqnarray}}

\addtocounter{section}{-1}
\begin{document}

\begin{center}

\vspace{2in}

{\large {\bf{Nonlinear optical response from excitons\\
in soliton lattice systems\\
of doped conjugated polymers
} } }

\vspace{1cm}

(Running head: {\sl Nonlinear optical response in doped conjugated
polymers})

\vspace{1cm}

{\rm Kikuo Harigaya\footnote[1]{E-mail address:
harigaya@etl.go.jp; URL: http://www.etl.go.jp/People/harigaya/.}}\\

\vspace{1cm}

{\sl Fundamental Physics Section,\\
Electrotechnical Laboratory,\\
Umezono 1-1-4, Tsukuba, Ibaraki 305, Japan}

\vspace{1cm}

(Received ~~~~~~~~~~~~~~~~~~~~~~~~~~~~~~~~~)
\end{center}

\vspace{1cm}

\noindent
{\bf Abstract}\\
Exciton effects on conjugated polymers are investigated
in soliton lattice states.  We use the Su-Schrieffer-Heeger
model with long-range Coulomb interactions.  The
Hartree-Fock (HF) approximation and the single-excitation
con\-fig\-u\-ra\-tion-interaction (single-CI) method are used
to obtain optical absorption spectra.  The third-harmonic
generation (THG) at off-resonant frequencies is calculated
as functions of the soliton concentration and the chain
length of the polymer.  The magnitude of the THG at the 10
percent doping increases by the factor about 10$^2$ from that
of the neutral system.  This is owing to the accumulation of
the oscillator strengths at the lowest exciton with increasing
the soliton concentration.  The increase by the order two is
common for several choices of Coulomb interaction strengths.

\mbox{}


\pagebreak

\section{Introduction}

Recently, effects of electron-electron interactions have been
investigated in many aspects of conjugated polymers.  For example,
electronic excitation structures in the half-filled conjugated
polymers with the constant bond alternation have been theoretically
investigated by using the exciton formalism [1] and the exact
diagonalization method [2], and also by solving the time-dependent
Hartree-Fock (HF) equations [3].  The roles of excitons were
pointed out earlier, but excitation structures have been considered
intensively still recently, relating with origins of nonlinear
optical spectra [2-4].  The lowest energy excitation has the
largest oscillator strength as the most remarkable consequence of
correlation effects.  This is clearly seen when the optical
spectra calculated by using the HF wavefunctions are compared
with the spectra with the correlation effects.  This feature is
observed when the higher correlations are taken into account
by the single-excitation configuration-interaction (single-CI)
method [1] and also by the time-dependent HF formalism [3].

It is widely known that the soliton, polaron, and bipolaron
lattices are present [5], when the SSH model [6], its continuum
version [7], and the extended model with the term of the nondegeneracy
[8] are doped with electrons or holes.  The new bands related with
the nonlinear excitations develop in the Peierls gap as the doping
proceeds.  When correlation effects are considered by the single-CI,
the excitation structures exhibit the presence of excitons.
In the previous paper [9], we have considered exciton effects
in the soliton lattice states of the doped conjugated polymers.
There is one kind of exciton in the half-filled system, where the
excited electron (hole) sits at the bottom of the conduction
band (top of the valence band).  We have called this exciton as the
``intercontinuum exciton" [9].  In the soliton lattice states of the
doped SSH model for degenerate conjugated polymers, there are
small gaps between the soliton band and the continuum states,
i.e., valence and conduction bands.  Therefore, the number of the
kind of excitons increases, and their presence is reflected in
structures of the optical spectra.  A new exciton, which we have
named the ``soliton-continuum exciton" [9], appears when the
electron-hole excitation is considered between the soliton band and
one of the continuum bands.  We have looked at variations of
relative oscillator strengths of the new excitons, the
soliton-continuum and intercontinuum excitons.  When the
excess-electron concentration is small, the ratio of the oscillator
strengths of the soliton-continuum exciton increases almost
linearly with respect to the concentration.  The oscillator
strengths accumulate rapidly at this exciton as the concentration
increases.  The contribution from the soliton-continuum exciton
is more than 90 percent at the 10\% doping.

The purpose of this paper is to investigate how the above changes
of the characters of optical excitations are reflected in the
nonlinear optical properties.  We might be able to realize large
optical nonlinealities in soliton lattice systems, because the
energy gap is small in doped conjugated polymers and the oscillator
strengths accumulate rapidly at the soliton-continuum exciton as the
doping proceeds.  Even though experimental facts of the doped polymers
have not been reported so much, it would be quite interesting to
demonstrate theoretically how large optical nonlinearities would be
obtained when conjugated polymers are doped with electrons
or holes up to as much as 10 percent.  The SSH model with
the long range Coulomb interactions of the Ohno expression [10]
is solved with the HF approximation, and the excitation
wavefunctions of electron-hole pairs are calculated by the
single-CI.   We consider the off-resonant nonlinear susceptibility
as a guideline of the magnitude of the nonlinearity.  In this
case, multi excitations, such as double (triple) excitations [11],
do not contribute significantly, so we can assume that the single-CI
well describes the characters of excitations at off-resonances.
We will calculate the third harmonic generation (THG)
$\chi_{\rm THG}^{(3)} (\omega) = \chi(3\omega;\omega,\omega,\omega)$
at the zero frequency $\omega = 0$, with changing the chain length
and the soliton concentration.  We will show that the magnitude
of the THG at the 10 percent doping increases by the factor about
10$^2$ from that of the neutral system.  This is owing to the
accumulation of the oscillator strengths at the lowest exciton with
increasing the soliton concentration.  The increase by the order
two is common for several choices of Coulomb interaction strengths.

This paper is composed as follows.  In \S 2, the model is
introduced and the numerical method is explained.  Results
of the optical nonlinearity are reported and discussion
is made in \S 3.  The paper is summarized in the final section.

\section{Model}

We use the SSH hamiltonian [6] with the Coulomb interactions:
\beeq
H = H_{\rm SSH} + H_{\rm int}.
\eneq
The first term of eq. (1) is:
\beeqa
H_{\rm SSH} &=& - \sum_{i,\sigma} ( t - \alpha y_i )
( c_{i,\sigma}^\dagger c_{i+1,\sigma} + {\rm h.c.} )  \nonumber \\
&+& \frac{K}{2} \sum_i y_i^2,
\eneqa
where $t$ is the hopping integral of the system without the
bond alternation; $\alpha$ is the electron-phonon coupling constant
which changes the hopping integral linearly with respect to the
bond variable $y_i$; $c_{i,\sigma}$ is an annihilation operator of
the $\pi$-electron at the site $i$ with spin $\sigma$; the sum is
taken over all the lattice sites of the periodic chain; and the
last term with the spring constant $K$ is the harmonic energy of
the classical spring simulating the $\sigma$-bond effects.  The
second term of eq. (1) is the long-range Coulomb interaction in
the form of the Ohno potential [10]:
\beeqa
H_{\rm int} &=& U \sum_i
(c_{i,\uparrow}^\dagger c_{i,\uparrow} - \frac{n_{\rm el}}{2})
(c_{i,\downarrow}^\dagger c_{i,\downarrow} - \frac{n_{\rm el}}{2}) \nonumber \\
&+& \sum_{i \neq j} W(r_{i,j})
(\sum_\sigma c_{i,\sigma}^\dagger c_{i,\sigma} - n_{\rm el})
(\sum_\tau c_{j,\tau}^\dagger c_{j,\tau} - n_{\rm el}),
\eneqa
where $n_{\rm el}$ is the number of $\pi$-electrons per site,
$r_{i,j}$ is the distance between the $i$th and $j$th sites, and
\beeq
W(r) = \frac{1}{\sqrt{(1/U)^2 + (r/a V)^2}}
\eneq
is the Ohno potential.  The quantity $W(0) = U$ is the strength of
the onsite interaction, $V$ means the strength of the long range part,
and $a$ is the mean bond length.

The model is treated by the HF approximation and the single-CI
for the Coulomb potential.  The adiabatic approximation is applied
to the bond variables.  The HF order parameters and bond variables
are determined selfconsistently using the iteration method [12].
The details of the formalism have been explained in the previous
paper [9].  A geometry of a ring with the coordinate of $j$th
carbon atom ($1 \leq j \leq N$),
\beeq
(R \cos \frac{2 \pi j}{N}, R \sin \frac{2 \pi j}{N}, 0),
\eneq
is used for a polymer chain, in order to remove edge effects.
Here, $R = Na/(2\pi)$ is the radius of the polymer ring, $N$ is
the system size, and $a$ is the lattice constant.  The electric
field of light is parallel to the $x$-$y$ plane.  The optical
absorption spectra where light is along with the $x$-
and $y$-directions are summed as shown by the following formula:
\beeq
\sum_\kappa E_{\kappa} P (\omega - E_{\kappa})
(\langle g | x |\kappa \rangle \langle \kappa | x | g \rangle
+ \langle g | y |\kappa \rangle \langle \kappa | y | g \rangle).
\eneq
Here, $P (\omega) = \gamma/[ \pi (\omega^2 + \gamma^2)]$
is the Lorentzian distribution ($\gamma$ is the width),
$E_{\kappa}$ is the electron-hole excitation energy,
$| \kappa \rangle$ is the $\kappa$th excitation, and
$| g \rangle$ means the ground state.

The THG is calculated with the conventional formula [13-15]:
\beeqa
\chi^{(3)} (3\omega;\omega,\omega,\omega) &=& e^4 N_{\rm d}
\sum_{\kappa,\lambda,\mu} f_{g,\mu} f_{\mu,\lambda} f_{\lambda,\kappa}
f_{\kappa,g} \nonumber \\
&\times& [ \frac{1}{(E_{\mu,g}-3\omega)
(E_{\lambda,g}-2\omega)(E_{\kappa,g}-\omega)} \nonumber \\
&+& \frac{1}{(E_{\mu,g}^*+\omega)
(E_{\lambda,g}-2\omega)(E_{\kappa,g}-\omega)} \nonumber \\
&+& \frac{1}{(E_{\mu,g}^*+\omega)
(E_{\lambda,g}^*+2\omega)(E_{\kappa,g}-\omega)} \nonumber \\
&+& \frac{1}{(E_{\mu,g}^*+\omega)
(E_{\lambda,g}^*+2\omega)(E_{\kappa,g}^*+3\omega)} ],
\eneqa
where, $N_{\rm d}$ is the number density of the crystalline polymer,
and this is material-dependent.  For demonstration of the magnitude
of the THG, we use the value of the number density of the CH unit,
which is taken from {\sl trans}-polyacetylene:
$N_{\rm d}=5.24\times 10^{22} {\rm cm}^{-3}$ [16].
We also use $t=1.8$eV in order to look at numerical data in the esu unit.
In eq. (7), $f_{\lambda,\kappa} = \langle \lambda | x | \kappa \rangle$
is the dipole matrix element with the electric field parallel
with the $x$ axis, $E_{\kappa,g} = E_\kappa - E_g$, $E_\kappa$ is
the energy of the excited state, and $E_g$ is the energy of
the ground state.  In the actual calculation, we change the order
of terms so as to take into account the mutual cancellation among
them [17].  Also, we include a small imaginary part in the denominator:
for example, $E_{\kappa,g} \rightarrow E_{\kappa,g} + {\rm i} \eta$ and
$E_{\kappa,g}^* \rightarrow E_{\kappa,g}^* - {\rm i} \eta$.
This assumes a lifetime broadening, and suppresses the height of
the delta-function peaks.  The THG at $\omega = 0$ does not
sensitively depend on the choice of $\eta$.  This can be
checked by varying the broadening.  In the next section, we report
the results with the value $\eta = 0.02 t$.

The system size is chosen as $N= 80, 100, 120$ when the electron
number is even (it is varied from $\nel =
N, N+2, N+4, N+6, N+8, N+10$ to $N+12$),
because the size around 100 is known to give well the energy gap
value of the infinite chain.  More larger system size becomes tedious
for doing single-CI calculations which call for huge computer
memories.  In principle, we have to adjust parameters and find appropriate
ones in order to reproduce experimental data, such as, the energy
gap and the bond alternation amplitude.  But, we will change parameters
arbitrary in a reasonable range in order to look at general properties
of the optical nonlinearity.  We take two combinations of the Coulomb
parameters $(U,V) = (2t,1t)$ and $(4t,2t)$ as the representative cases.
The other parameters, $t = 1.8$eV, $K = 21$eV/\AA$^2$, and
$\alpha = 4.1$eV/\AA, are fixed in view of the general interests
of this paper.  All the quantities of energy dimension are shown
in the units of $t$.

\section{Numerical Results and Discussion}

Figure 1(a) shows the optical absorption spectrum at the 2\%
soliton concentration for $(U,V) = (4t,2t)$.  The broadening
$\gamma = 0.05t$ is used.  The same data have been used in
Fig. 3(b) of the ref. [9].  There are two main features around
the energies 0.7$t$ and 1.4$t$.  The former originates from
the soliton-continuum exciton, and the latter is from the
intercontinuum exciton.

Figure 1(b) displays the absolute value of the THG against the
excitation energy $\omega$.  The abscissa is scaled by the
factor 3 so that the features in the THG locate at the similar
points in the abscissa of Fig. 1(a).  The small feature at about
$\omega=0.22t$ comes from the lowest excitation of the
soliton-continuum exciton and the larger features at about
$\omega=0.24t$ and $0.32t$ come from the higher excitations.
The features from the intercontinuum exciton extend from
$\omega = 0.48t$ to the higher energies.  In the present
calculations, the THG in the energy region higher than 0.5$t$
is not large relatively.  The point, that the THG becomes
smaller as the excitation energy increases, has been seen
in the calculations of the half-filled system [4].  However,
in the time-dependent HF formalism, the THG in higher energies
is still larger as shown by Fig. 4 of the ref. [3].  The
difference of the distribution of the THG strengths might
come from the difference of the approximation method for
electron correlations.

The THG data like in Fig. 1(b) are calculated for the three system
sizes, $N=80$, 100, 120, and for the soliton concentrations
up to 10\%.  As clearly seen for example in Fig. 1(b), the
off-resonant THG at $\omega = 0$ is quite far from features
coming from excitons.  The contributions from double (triple and so on)
excitations would be very small.  Thus, the single-CI calculations
could be used as a measure of the optical nonlinearities of
doped conjugated polymers.

Figures 2(a) and (b) display the variations of the absolute
value of $\chi_{\rm THG}^{(3)} (0)$ for $(U,V) = (2t,1t)$
and $(4t,2t)$, respectively.  The plots are the numerical data,
and the dashed lines are the guide for eyes, showing the overall
behavior for each system size.  The deviations of the plots
from the expected smooth behavior might come from the quantum
effect due to the finite system size [18].  The linear absorption
has the size consistensy, so the plots of the relative oscillator
strengths vary smoothly as functions of the
soliton concentration [9].  However, the THG is not size
consistent, and spectral shapes depend on the system size when
$N$ is as large as 100 [18].  Therefore, it would not be
strange even if $|\chi_{\rm THG}^{(3)}| (0)$ is sensitive to
the system size and the soliton concentration.  The THG
increases as the system size increases.  This behavior is
the same as has been seen in the calculations of the half-filled
systems [18].

The increase of the off-resonant THG near zero concentration is very rapid,
but the THG is still increasing for a few percent to 10\%
soliton concentration.  The behavior is similar for the two sets
of the Coulomb interactions for Figs. 2(a) and (b).  The THG in
Fig. 2(b) is about ten times smaller than that of Fig. 2(a).
This is due to the difference of the interaction strengths.
The increase by about the factor 10$^2$ at the 10\% doping
from that of the neutral system seems to be a remarkable fact.
This fact might be independent of Coulomb interactions strengths
characteristic to optical excitation of the system.  The
experiments of nonlinear optics on doped conjugated polymers
have not been reported so many times.  However, the present
theory could used as one of guidelines for strengths of
optical nonlinearities in doped systems.

Then, why such the large increase of the THG would occur upon
doping of the polymers?  In the soliton lattice
theory with the continuum model [5],
the energy gap decreases as the soliton concentration increases.
Therefore, it may seem first that the decrease of the energy gap
is one of the reasons.  But, as shown in the Fig. 4 of the ref. [9],
the lowest optical gap due to the single-CI calculations
is almost independent of the concentration,
and thus the change of the optical gap would not be the main
reason.  However, as we have discussed in [9], the ratio of the
oscillator strength of the soliton-continuum exciton increases
very rapidly.  In view of this change of the exciton characters,
it would be natural to conclude that the increase of the THG
by the factor 10$^2$ is due to the fact that the oscillator
strengths accumulate rapidly at the lowest exciton with increasing
the soliton concentration.

\section{Summary}

We have considered the off-resonant nonlinear susceptibility
as a guideline of the strength of the nonlinearity in the doped
conjugated polymers.   We have calculated the off-resonant
THG with changing the chain length and the soliton concentration.
We have shown that the magnitude of the THG at the 10 percent
doping increases by the factor about 10$^2$ from that of the
neutral system.  The increase by the order two is common for
the several choices of Coulomb interaction strengths.

\mbox{}

\noindent
{\bf Acknowledgements}\\
The author acknowledges useful discussion with
Prof. T. Kobayashi, Prof. S. Stafstr\"{o}m, Dr. S. Abe,
Dr. Y. Shimoi, and Dr. Akira Takahashi.

\pagebreak
\begin{flushleft}
{\bf References}
\end{flushleft}

\noindent
$[1]$ Abe S, Yu J and Su W P 1992 {\sl Phys. Rev.} B {\bf 45} 8264\\
$[2]$ Guo D, Mazumdar S, Dixit S N, Kajzar F,
Jarka F, Kawabe Y and Peyghambarian N 1993 {\sl Phys. Rev.}
B {\bf 48} 1433\\
$[3]$ Takahashi A and Mukamel S 1994 {\sl J. Chem. Phys.}
{\bf 100} 2366\\
$[4]$ Abe S, Schreiber M, Su W P and Yu J 1992
{\sl Phys. Rev.} B {\bf 45} 9432\\
$[5]$ Horovitz B 1981 {\sl Phys. Rev. Lett.} {\bf 46} 742\\
$[6]$ Su W P, Schrieffer J R and Heeger A J 1980
{\sl Phys. Rev.} B {\bf 22} 2099\\
$[7]$ Takayama H, Lin-Liu Y R, and Maki K
1980 {\sl Phys. Rev.} B {\bf 21} 2388\\
$[8]$ Brazovskii S A and Kirova N N 1981
{\sl JETP Lett.} {\bf 33} 4\\
$[9]$ Harigaya K, Shimoi Y and Abe S 1995
{\sl J. Phys.: Condens. Matter} {\bf 7} to be published\\
$[10]$ Ohno K 1964 {\sl Theor. Chem. Acta} {\bf 2} 219\\
$[11]$ Shakin V A and Abe S {\sl Phys. Rev.} B {\bf 50} 4306\\
$[12]$ Terai A and Ono Y 1986 {\sl J. Phys. Soc. Jpn.}
{\bf 55} 213\\
$[13]$ Bloembergen N {\sl Nonlinear Optics} (Benjamin, New York, 1965)\\
$[14]$ Orr B J and Ward J F 1971 {\sl Mol. Phys.} {\bf 20} 513\\
$[15]$ Yu J, Friedman B, Baldwin P R and Su W P 1989
{\sl Phys. Rev.} B {\bf 39} 12814\\
$[16]$ Fincher C R, Chen C E, Heeger A J, MacDiarmid A G
and Hastings J B 1982 {\sl Phys. Rev. Lett.} {\bf 48} 100\\
$[17]$ Yu J and Su W P 1991 {\sl Phys. Rev.} B {\bf 44} 13315\\
$[18]$ Abe S, Schreiber M, Su W P and Yu J 1992 {\sl J. Lumin.}
{\bf 53} 519\\

\pagebreak

\begin{flushleft}
{\bf Figure Captions}
\end{flushleft}

\mbox{}

\noindent
Fig. 1. (a) The optical absorption spectrum and (b) the absolute
value of the THG, for the system size $N=100$, the electron
number $\nel = 102$, and $(U,V) = (4t,2t)$.  The broadening
$\gamma = 0.05t$ is used in (a), and $\eta = 0.02t$ is used
in (b).
\mbox{}

\mbox{}

\noindent
Fig. 2.  The absolute value of the THG at $\omega = 0$ v.s.
the soliton concentration for (a) $(U,V) = (2t,1t)$ and (b) $(4t,2t)$.
The numerical data are shown by the triangles ($N=80$),
circles ($N=100$), and squares ($N=120$), respectively.
The dashed lines are the guide for eyes.

\end{document}